\documentclass[reqno]{amsart}
\usepackage[top=0.5in,bottom=0.5in,left=0.5in,right=0.5in]{geometry}
\usepackage{amscd}

\usepackage{multicol}
\usepackage{times}
\usepackage[T1]{fontenc}
\usepackage[latin1]{inputenc}

\usepackage[arrow,curve,matrix]{xy}

\usepackage{graphicx}
\usepackage{extarrows}
\usepackage{flafter}
\usepackage{placeins}
\usepackage{enumitem}
\usepackage[OT2,T1]{fontenc}
\DeclareSymbolFont{cyrletters}{OT2}{wncyr}{m}{n}
\DeclareMathSymbol{\Sha}{\mathalpha}{cyrletters}{"58}

\usepackage{centernot}

\usepackage{mathrsfs}
\usepackage[color]{showkeys}
\usepackage{url}

\definecolor{refkey}{rgb}{1,1,1}
\definecolor{labelkey}{rgb}{1,1,1}
\definecolor{cite}{rgb}{0.9451,0.2706,0.4941}
\definecolor{ruri}{rgb}{0.0078,0.4022,0.8010}

\usepackage[%
bookmarks=true,bookmarksnumbered=true,%
colorlinks=true,linkcolor=ruri,citecolor=red%
 ]{hyperref}

\def\P{{\rm \mathbb{P}}}

\theoremstyle{plain}

\newtheorem{theorem}{Theorem}[section]

\newtheorem{proposition/example}[theorem]{Proposition/Example}

\theoremstyle{definition}

\newtheorem{example}[theorem]{Example}

\newtheorem{conjecture/question}[theorem]{Conjecture/Question}

\newtheorem{remark/definition}[theorem]{Remark/Definition}
\newtheorem{definition/notation}[theorem]{Definition/Notation}

 \theoremstyle{remark}

\numberwithin{equation}{section}

\usepackage{extarrows}\usepackage{bm}

\begin{document}
\title {Universal Constants, Law of Inertia and Emergent Geometry}

\author{Zhi Hu}

\address{ \textsc{School of Mathematics, Nanjing University of Science and Technology, Nanjing 210094, China}}

\email{halfask@mail.ustc.edu.cn}
\author{Mulin Yan}

\address{School of Mathematics, University of Science and
Technology of China\\Hefei, 230026, China}
 \email{
 mlyan@ustc.edu.cn }
\author{Runhong Zong}

\address{ \textsc{Department of Mathematics, Nanjing University, Nanjing 210094, China}}

\email{rzong@nju.edu.cn}

\maketitle
\begin{abstract}
In this paper, we only treat the law of inertia as the first principle, then a nontrivial geometry emerges by introducing more universal constants,  in which the main ideas appearing in  deformed special relativity (DSR), (Anti-)de Sitter special relativity [(A)dSSR] and bimetric gravity (BMG) have been contexture.
\end{abstract}

\section{Introduction}
 The de Sitter and  anti-de Sitter spacetimes are  of the most symmetric solutions
of Einstein's field equations including the cosmological constant. For this reason,
they are  important for general relativity. After 1998, these spacetimes have drawn attention of high energy physicists due to the
conjectured (anti-)de Sitter space/conformal field theory [(A)dS/CFT] correspondence.
In this letter, we propose a new mechanism to produce  de Sitter and  anti-de Sitter spacetimes  from the law of inertia of massive free particles.

 Our initial motivation is to consider a fundamental   theory of relativity that admits more universal constants.
In Galilei relativity there is no observer-independent scale, and
Einstein's special
relativity (SR) introduced the first observer-independent relativistic scale: the velocity scale $c$ identified
with the speed of light. Naturally, the second  observer-independent scale could be considered as length. It is clear that there is
an  inevitable price to pay for  admitting at the same time
relativity principle and the observer-independent scale of length.
That is, analogous to  the Galilei$\rightarrow$Einstein transition, one should  deform Poincar\'{e} group which has already deformed Galilean  group through the contraction limit of infinity $c$.

There are two possible scenarios:  in order to describe ultra-short-distance or ultra-large-distance  physics, we might have to set aside
SR and replace it with a new relativity theory with two characteristic invariant
scales.
The extra universal constant with dimension of length is denoted by $\ell$.

-- $\ell$ is very small identified with the Planck length $\sim 10^{-35}m$, therefore the modified theory, called deformed special relativity or doubly special relativity (DSR),  may be rooted in quantum gravity \cite{21,112,le,lee}. The best
developed approach to DSR is realized  based on the so-called $\ell$-Poincar\'{e} algebra and
$\ell$-Minkowski spacetime. Here, the  usual Poincar\'{e} algebra has been deformed into a quantum Hopf algebra which can be understood as
the symmetry algebra of a noncommutative deformation
of usual Minkowski spacetime \cite{jk,jkk}.   In the  low-energy limit,  i.e. $\ell$ tending to zero, everything returns to the standard SR.

-- $\ell$ is very large identified with the radius   of  (observable) universe $\sim 10^{26}m$, then things could turn into relatively easy. At hand, we have the (Anti)-de Sitter group,  which is interpreted as a particular deformation of
the Poincar\'{e} group through the contraction
limit of infinity $\ell$. Hence, conceptually, one should establish special  relativity with invariance under (Anti)-de Sitter  group [(A)dSSR]. To our knowledge, this  theory was first  suggested by Dyson in his  famous paper
\cite{dy} and independently by the authors of \cite{5}, and was was  further developed in \cite{51,52,53,54,55,56,57}.

-- The above two scenarios can be combined together to construct an extension of SR characterized by three invariant scales: in addition to $c$,  two universal constants $\ell_1,\ell_2$ with dimensions of length are included, which are identifies  with  Planck length and radius   of   universe, respectively. Such  theory will reduce to DSR when $\ell_2$ goes to infinity and reduce to (A)dSSR when $\ell_1$ tends to zero. A proposal   called triply special relativity has been  described  by a new nonlinear deformation of Poincar\'{e} algebra in \cite{sm}.

Our method  is very different. More precisely,
we start with the usual  spacetime $\mathbb{R}^4$ equipped with the Minkowski metric  $(\eta_{\mu\nu})=\textrm{diag}(-1,1,1,1)$, then  we allow extra universal constant $\ell$  to appear in the dynamical part of theory. If the background is fixed to be Minkowski spacetime without any modification, it seems that it is  the unique approach to introduce the new universal constants.  We will show  a nontrivial  geometry emerges from the dynamical structure. For simplicity, we only work within the single particle sector of the theory. Analogous to the standard action for a free particle with mass $m$ in SR
\begin{align}
  S=-m\int dt \sqrt { |\eta_{\mu\nu} v^\mu v^\nu|}.
\end{align}
where   $c=1$ is set, and $v^\mu=\frac{dx^\mu}{dt}$, we write
a new action
\begin{align}\label{0}
   S=-m\int dt \sqrt { |B_{\mu\nu} v^\mu v^\nu|},
 \end{align}
where the symmetric second-order tensor $B_{\mu\nu}$
is not chosen  a priori to be equal to $\eta_{\mu\nu}$.
In other words,   the geometry of the background  and the dynamics of matter  are separately considered, and  described by two different (but maybe relevant) symmetric second-order tensors $\eta_{\mu\nu}$ and $B_{\mu\nu}$, respectively.
Actually, this idea is similar with  the theory of  bimetric gravity (BMG). That theory also
consists of two metric-like  symmetric second-order tensors which play different roles \cite{tu}. The first one is surely the dynamical metric that describe the geometry of spacetime and thus the gravitational field,  and the second one can be non-dynamical or dynamical. For example, in Rosen-type theory,  the  second metric refers to the Minkowski metric  and describes the inertial forces \cite{rn}, and in Hassan-Rosen-type theory which is free from the Boulware-Deser ghost and propagates seven degrees of freedom, the introduction of the second metric nonlinearly  coupled to the spacetime metric allows for
a description of a massive spin-2 field \cite{r,rr,rrr}.

Coming back to our theory, now $B_{\mu\nu}$ does not arise from background geometry any more, but is still  tied down by dynamics.  Specifically speaking, the dynamical content of the massive free particle in Minkowski spacetime is just the  law of inertia, hence,  if  the  action \eqref{0} correctly produces  the dynamics of massive free particle  according to the least action  principle,   the law of inertia will continue to hold true. This provides a constraint on the form of $B_{\mu\nu}$.
We will see that under some suitable domain in Minkowski spacetime, $B_{\mu\nu}$ can be exactly viewed as a ``metric"  on  the maximally symmetric spacetime  with nonzero curvature (cosmological constant), i.e. (A)dS spacetime such that at the level of practice, this dynamics is equivalent to (A)dSSR. In this sense, we say that a nontrivial geometry [i.e. (A)dS geometry (or (A)dSSR) in the present case] emerges from the dynamics of massive free particles -- the law of inertia. We can call such emergent metric the inertial metric, analogous to inertial force.

It is noteworthy that
 the above process is only valid for the  large scale $\ell$, namely the theory makes no sense if $\ell$ goes to zero. On the other hand,  in principle, the choice of our action for  a massive free particle has a lot of freedom, as long as it produces the right law of inertia. This reveals that considering more general action would allow us to introduce more  universal constants, in particular, those involved the  small scale. Of course, some dynamical symmetries will disappear.

An effortless manner is using the  pair $(\eta_{\mu\nu},B_{\mu\nu})$ to construct the  following  bipartite-Finsler-like action
\begin{align}\label{41}
   S=-m\int dt (\sqrt { |\eta_{\mu\nu} v^\mu v^\nu|}+\xi \sqrt { |B_{\mu\nu} v^\mu v^\nu|}).
 \end{align}
 Now $\xi$ is a dimensionless constant such that to kill the dimension we can simultaneity introduce more universal constants with the same dimension. For example,  we can  puts $\xi=\frac{\ell_1}{\ell_2}$, where $\ell_1,\ell_2$ are the universal constants identified with  Planck length and  radius of universe,  respectively. Now the theory returns to the usual one when  $\ell_1$ tend to zero or $\ell_2$ tends to infinity.   If one  picks $B_{\mu\nu}$ as the inertial metric mentioned previously, it is obvious that  this action can also describe the massive free particle in physical domains in the sense of preserving the law of inertia.
 The emergent geometry from the action \eqref{41} can be described by  the Finsler metric, which is exactly   the second-order derivative of the corresponding Lagrangian with respect to the 4-velocity.
 Then the   dynamical effects can be studied under the framework of Finsler geometry \cite{baa}. It's also worth mentioning that  the action of type \eqref{41} has been used to investigate the  Lorentz violation  in \cite{z1,z2,z3}, and  Finsler geometry also provides  a geometric tool in some modified special relativity theories (DSR \cite{ggg,mi,gf}, very special relativity (VSR) \cite{gg,gg1}).

In conclusion, we only treat the law of inertia as the first principle, then a nontrivial geometry emerges by introducing more universal constants,  in which the main ideas appearing in  DSR, (A)dSSR and BMG have been contexture.

 \section{Solutions of Law of Inertia with Two Universal Constants}
    We need to determine the general form of $B_{\mu\nu}$ based on some reasonable assumptions. Firstly, according to our target, $B_{\mu\nu}$ should
   involve the universal constants $c$ and $\ell$. Secondly, when $\ell$ tends to infinity or zero, $B_{\mu\nu}$ should revert to  $\eta_{\mu\nu}$.
  Then,  taking into account the dimension, we  pick the following very general ansatz
\begin{align}\label{nj}
  B_{\mu\nu}=A_0\eta_{\mu\nu}+\sum_{I=1}^dA_{I}\frac{(x\cdot v)^{a_I-2}(v\cdot v)^{b_I}\eta_{\mu\alpha}\eta_{\nu\beta}x^\alpha x^\beta}{\ell^{a_I}},
\end{align}
where for the  two 4-vectors $\Theta^\mu,\Xi^\nu$, one defines $$\Theta\cdot \Xi=\eta_{\mu\alpha}\Theta^\alpha \Xi^\mu,$$
and the  integers $a_I$, $b_I$ satisfy
\begin{itemize}
  \item $a_I+2b_I=2$,
  \item $a_I\neq 0$,
  \item all $a^I$ have the same sign.
\end{itemize}
 The above action  is recognized to be Finsler-like, and the corresponding Finsler function \cite{baa} is exactly the Lagrangian given by
 \begin{align}\label{1}
 &L(t,x^i,v^i)\nonumber\\
 =&\ -m\sqrt{|A_0v\cdot v+\sum_{I=1}^dA_I(x\cdot v)^{a_I}(v\cdot v)^{b_I}|},
\end{align}
 where as before, we put  $\ell=1$  for convenience.

   Note that $B_{\mu\nu}$ generally dose not only depend on the coordinates on   $\mathbb{R}^4$, even is  not necessary to be well-defined over the entire $\mathbb{R}^4$.  The domain lying $\mathbb{R}^4$ such that $B_{\mu\nu}$ makes sense (for example, well-defined, non-degenerate and has  the suitable  signature) is called the physical domain.

If a massive free particle  is assumed to be subject to our new  action \eqref{0}, then the corresponding  Euler-Lagrange equation  should imply the the law of inertia, that is the acceleration of particle has to vanish.
The Euler-Lagrange equation reads
\begin{align}
  \frac{\partial L}{\partial x^i}=\frac{\partial^2 L}{\partial t\partial v^i}+v^j\frac{\partial^2 L}{\partial x^j\partial v^i}+\frac{\mathrm{d}v^j}{\mathrm{d}t}\frac{\partial^2 L}{\partial v^j v^i},
\end{align}
therefore we must have
 \begin{align}\label{2}
       \frac{\partial L}{\partial x^i}-\frac{\partial^2 L}{\partial t\partial v^i}-v^j\frac{\partial^2 L}{\partial x^j\partial v^i}&=0,\\
\det(\frac{\partial^2 L}{\partial v^j\partial v^i})&\neq0.
 \end{align}

Substituting  the expression \eqref{1} into the equation \eqref{2} gives rises to
\begin{align}\label{sss}
&\ 2[\partial_iA_0(v\cdot v)+\sum_{I=1}^d\partial_iA_I(x\cdot v)^{a_I}(v\cdot v)^{b_I}][A_0(v\cdot v)+\sum_{I=1}^dA_I(x\cdot v)^{a_I}(v\cdot v)^{b_I}]\nonumber\\
=&\ 2[2(v\cdot \partial A_0)v^i+\sum_{I=1}^d(v\cdot\partial A_I)a_I(x\cdot v)^{a_I-1}(v\cdot v)^{b_I}x_i\nonumber\\
&\ \ \ +\sum_{I=1}^d A_Ia_I(a_I-1)(x\cdot v)^{a_I-2}(v\cdot v)^{b_I+1}x_i +2\sum_{I=1}^d (v\cdot\partial A_I)b_I(x\cdot v)^{a_I}(v\cdot v)^{b_I-1}v_i\nonumber\\
&\ \ \ +2\sum_{I=1}^d A_Ia_Ib_I(x\cdot v)^{a_I-1}(v\cdot v)^{b_I}v_i][A_0(v\cdot v)+\sum_{I=1}^dA_I(x\cdot v)^{a_I}(v\cdot v)^{b_I}]\nonumber\\
&-[2A_0v^i+\sum_{I=1}^dA_Ia_I(x\cdot v)^{a_I-1}(v\cdot v)^{b_I}x_i+2\sum_{I=1}^d A_Ib_I(x\cdot v)^{a_I}(v\cdot v)^{b_I-1}v_i]\nonumber\\
&\times[(v\cdot\partial A_0)(v\cdot v)+\sum_{I=1}^d(v\cdot\partial A_I)(x\cdot v)^{a_I}(v\cdot v)^{b_I}+\sum_{I=1}^d A_Ia_I(x\cdot v)^{a_I-1}(v\cdot v)^{b_I+1}],
\end{align}
where the
following notations are employed
\begin{align*}
       \partial_i A&=\frac{\partial A}{\partial x^i},\\
        \partial A&=(-\frac{\partial A}{\partial t}, \frac{\partial A}{\partial x^1},\frac{\partial A}{\partial x^2}, \frac{\partial A}{\partial x^3}),
      \end{align*}
and $\Xi_\mu=\eta_{\mu\alpha}\Xi^\alpha$ for a 4-vector $\Xi^\mu$.
Comparing the monomials of the both sides of the equation \eqref{sss} with the same type, we find that only one index $I$ can survive such that   \eqref{sss} is simplified to the  following equations on $A_0,A_1$ with $a_1=2,b_1=0$:
  \begin{align*}
    \partial_i A_0&=2A_1x_i,\\
    (2A_1^2x_i+\partial_iA_1A_0)(x\cdot v)&=[2(v\cdot\partial A_1)A_0 -(v\cdot\partial A_0)A_1]x_i,\\
    \partial_iA_1(x\cdot v)&=(v\cdot \partial A_1)x_i,\\
    2(v\cdot\partial A_0)A_1&=4A_1^2(x\cdot v)=(v\cdot \partial A_1)A_0.
  \end{align*}
  These equations leads to
  \begin{align}
   \partial A_0&=2A_1x,\label{s}\\
\partial A_1A_0&=4A_1^2x.\label{ss}
  \end{align}
whose general solutions   are given by
 \begin{align}
  A_0&=\frac{A}{B+C\eta_{\mu\nu}x^\mu x^\nu},\\
  A_1&=-\frac{AC}{(B+C\eta_{\mu\nu}x^\mu x^\nu)^2}
\end{align}
for constants $A,B,C$. Consequently,  we obtain
\begin{align}\label{33}
  B_{\mu\nu}=&\ \frac{A}{B+C\frac{\eta_{\gamma\delta}x^\gamma x^\delta}{\ell^2}}\eta_{\mu\nu}-\frac{AC}{(B+C\frac{\eta_{\gamma\delta}x^\gamma x^\delta}{\ell^2})^2}\frac{\eta_{\mu\alpha}\eta_{\nu\beta}x^\alpha x^\beta}{\ell^{2}},
\end{align}
 where the universal constant $\ell$ is restored.

 A useful observation is  that  $B_{\mu\nu}$ can be written in terms of  the combination of two projection operators as
\begin{align}\label{333}
  B_{\mu\nu}=\frac{A}{B+C\frac{\eta_{\mu\nu}x^\mu x^\nu}{\ell^2}}\Phi_{\mu\nu}+\frac{AB}{(B+C\frac{\eta_{\mu\nu}x^\mu x^\nu}{\ell^2})^2}\Psi_{\mu\nu},
\end{align}
where
\begin{align}
 \Phi_{\mu\nu}&=\eta_{\mu\nu}-\frac{x_\mu x_\nu}{\eta_{\alpha\beta}x^\alpha x^\beta },\\
\Psi_{\mu\nu}&=\frac{x_\mu x_\nu}{\eta_{\alpha\beta}x^\alpha x^\beta }
\end{align}
satisfy the projection relations
\begin{align}
  \eta^{\nu\alpha}\Phi_{\mu\nu}\Phi_{\alpha\beta}&=\Phi_{\mu\beta},\\
  \eta^{\nu\alpha}\Psi_{\mu\nu}\Psi_{\alpha\beta}&=\Psi_{\mu\beta},\\
   \eta^{\nu\alpha}\Phi_{\mu\nu}\Psi_{\alpha\beta}&=0.\label{344}
\end{align}

 Since it is required that $B_{\mu\nu}$ is non-degenerate and it  tends to $\eta_{\mu\nu}$ up to an  insignificant constant conformal scalars when $\ell$ goes to infinity, we have
 \begin{itemize}
   \item $A\neq0, B\neq 0$,
   \item $AB>0$.
 \end{itemize}
To check the  condition  \eqref{2}, we only need to show that under the limit $l\rightarrow\infty$, which is straightforward calculated as
\begin{align*}
  \lim_{\ell\rightarrow\infty}\det(\frac{\partial^2 L}{\partial v^j\partial v^i})=m^3\frac{(\frac{A}{B})^{\frac{3}{2}}}{(|v\cdot v|)^{\frac{5}{2}}}\neq0.
\end{align*}
Obviously, if $C=0$  everything essentially goes back to the classical theory with  $B_{\mu\nu}=\eta_{\mu\nu}$.  Therefore, we consider $C\neq 0$, and it can be   assumed to be 1.

  To determined the  signature of  $B_{\mu\nu}$, we need to consider the signs of
\begin{align*}
  B_{00}&=-\frac{A(B+\bm{x}\cdot\bm{x})}{(B-t^2+\bm{x}\cdot\bm{x})^2}, \\
\widetilde{B}_{11}&=\frac{A(B+(x^2)^2+(x^3)^2)}{(B- t^2+\bm{x}\cdot\bm{x})(B+\bm{x}\cdot\bm{x})},\\
\det&\left(
      \begin{array}{cc}
        \widetilde{B}_{11} & \widetilde{B}_{12} \\
        \widetilde{B}_{12} & \widetilde{B}_{22} \\
      \end{array}
    \right)\\
    &=\frac{A^2(B+(x^3)^2)}{(B- t^2+ \bm{x}\cdot\bm{x})^2(B+\bm{x}\cdot\bm{x})},\\
    \det&\left(
      \begin{array}{ccc}
        \widetilde{B}_{11} & \widetilde{B}_{12} & \widetilde{B}_{13}\\
        \widetilde{B}_{12} & \widetilde{B}_{22} & \widetilde{B}_{23}\\
       \widetilde{B}_{13}& \widetilde{B}_{23}& \widetilde{B}_{33}\\
      \end{array}
    \right)\\
    &=
 \frac{A^3B}{(B- t^2+\bm{x}\cdot\bm{x})^3(B+\bm{x}\cdot\bm{x})},
                                                                                 \end{align*}
where $\widetilde{B}_{ij}=B_{ij}-\frac{B_{0i}B_{0j}}{B_{00}}$, $\bm{x}\cdot\bm{x}=(x^1)^2+(x^2)^2+(x^3)^2$. One easily finds

\begin{table}[ht]\caption{Signature of $B_{\mu\nu}$}\label{l}
  \begin{tabular}{c|c}
  \hline
  Condition & Signature \\
  \hline
  $A>0, B>0, B-t^2+\bm{x}\cdot\bm{x}>0$ & $(1,3)$  \\
  $A>0, B>0,  B-t^2+\bm{x}\cdot\bm{x}<0$ & $(4,0)$\\
  $A<0, B<0, B+\bm{x}\cdot\bm{x}<0$  & $(1,3)$\\
  $A<0, B<0, B-t^2+\bm{x}\cdot\bm{x}>0$ & $(2,2)$\\
   $A<0, B<0, B-t^2+\bm{x}\cdot\bm{x}<0, B+\bm{x}\cdot\bm{x}>0$ & $(1,3)$\\
  \hline
\end{tabular},
\end{table}

\noindent where  for  a nondegenrate symmetric matrix, we denote its signature by $(n_-,n_+)$ if it has $n_-$ negative eigenvalues  and $n_+$ positive eigenvalues, respectively.
In particular,  if $B_{\mu\nu}$ is required to have  the  the signature  $(1,3)$ as a spacetime-metric, we pick the physical domains
 \begin{itemize}
   \item (I): $B-t^2+\bm{x}\cdot\bm{x}>0$ for $A>0, B>0$ \label{m},
   \item (II): $B-t^2+\bm{x}\cdot\bm{x}<0$ for $A<0, B<0$ \label{mm}.
 \end{itemize}

\section{Emergent  $(A)dS_4$ Geometry }
Now we explain why $(A)dS_4$ geometry emerges from the above framework.   It is known that $(A)dS_4$
is defined by a  hypersurface in 5-dimensional  space $\mathbb{R}^5$ with the Minkowski  metric $\eta^{(5)}=\textrm{diag}(-1,1,1,1,1)
$ (or the metric $\widetilde{\eta}^{(5)}=\textrm{diag}(-1,-1,1,1,1)
$) via  the following equation \cite{ll}
\begin{align*}
    -T^2+X^2+Y^2+Z^2+bW^2=1 ( b>0),
\end{align*}
or\begin{align*}
  -T^2-bW^2+X^2+Y^2+Z^2=-1 ( b>0).
\end{align*}
Define the following coordinates which cover the half domain $\{W>0\}$ or $\{W<0\}$ in $(A)dS_4$
\begin{align}\label{t}
  x^0=\frac{T}{W},x^1=\frac{X}{W}, x^2=\frac{Y}{W},x^3=\frac{Z}{W},
\end{align}
Then the induced metric from $-dT^2+dX^2+dY^2+dZ^2\pm b dW^2 $on  this hypersurface  is given by in terms of the coordinate system $\{x^0,x^1,x^2,x^3\}$
\begin{align}\label{c}
  g_{\mu\nu}=\frac{\eta_{\mu\nu}}{b+\eta_{\alpha\beta}x^\alpha x^\beta}-\frac{\eta_{\mu\alpha}\eta_{\nu\beta}x^\alpha x^\beta}{(b+\eta_{\alpha\beta}x^\alpha x^\beta)^2},
\end{align}
or\begin{align}
  g'_{\mu\nu}=\frac{\eta_{\mu\nu}}{b-\eta_{\alpha\beta}x^\alpha x^\beta}+\frac{\eta_{\mu\alpha}\eta_{\nu\beta}x^\alpha x^\beta}{(b-\eta_{\alpha\beta}x^\alpha x^\beta)^2}.
\end{align}
They exactly coincide with our  $B_{\mu\nu}$  over physical domains (I) and (II), respectively,  up to   constant conformal scalars, in other words, the physical domains equipped with $B_{\mu\nu}$ can be viewed as the model of $(A)dS_4$-geometry. In some literature \cite{52,55},  $B_{\mu\nu}$ is called the Beltrami metric for $(A)dS_4$-geometry.

From this viewpoint, we
immediately conclude that the coordinate transformations preserve $B_{\mu\nu}$ form the group $O(1,4)$ or $O(2,3)$.
By contrast, this group is a dynamical symmetry group other than  geometric symmetry group as in (A)dSSR. For our theory,  Poinc\'{a}re group $ISO(1,3)$ is still the geometric symmetry group, and the overlap of these two classes of  symmetry groups is exactly  Lorentz group $O(1,3)$. Then  we can call Lorentz group the  inertial group since it consists of transformations  preserving the inertial motions in physical domain.  By decomposing a matrix belongs to the group $O(1,4)$ or $O(2,3)$ as
\begin{align*}
  \lambda\left(
    \begin{array}{cc}
      N& P\\
      \mp\frac{P^T\eta N}{\sqrt{1\mp\eta_{\mu\nu}P^\mu P^\nu}} & \sqrt{1\mp\eta_{\mu\nu}P^\mu P^\nu}\\
    \end{array}
  \right),
\end{align*}
with matrices $N=({N^\mu}_{\nu})$ and $P=(P^0,P^1,P^2,P^3)^T$ satisfing the relation
\begin{align}
                                                                          N^T\eta N=\eta+\frac{N^T\eta PP^T\eta N}{\mp1+\eta_{\mu\nu}P^\mu P^\nu},
                                                                         \end{align}
                                                                         where $\mp$ correspond $O(1,4)$ and  $O(2,3)$ respectively, and $\lambda$ is fixed to 1 or $-$1, then we can explicitly write these coordinate transformations  as fractional linear transformations
\begin{align}\label{y,ab,aa}
 x^\mu\mapsto\frac{{N^\mu}_{\nu} x^\nu+\sqrt{b}P^\mu}{\mp\frac{\eta_{\alpha\beta}{N^{\beta}}_{\gamma}P^\alpha x^\gamma}{\sqrt{1\mp\eta_{\mu\nu}P^\mu P^\nu}}+\sqrt{b}\sqrt{1\mp\eta_{\mu\nu}P^\mu P^\nu}},
\end{align}
 which come back to Poinc\'{a}re transformations when  $\ell$ tends to infinity.
By Norther method, we can easily obtain the   corresponding ten conserved charges
for a massive free particle \cite{53,55}.

By symmetry breaking, we can also construct the some other actions with less symmetries, which are closed related to the violation of the law of inertia.
The Lie bracket among the basis $\{M_{AB}=-M_{BA}, A,B=0,\cdots,4\}$ of Lie algebra $\mathfrak{o}(1,4)$ of de Sitter group $O(1,4)$ is given by
\begin{align*}
  [M_{AB},M_{CD}]=\eta^{(5)}_{AD}M_{BC}+\eta^{(5)}_{BC}M_{AD}-\eta^{(5)}_{AC}M_{BD}-\eta^{(5)}_{BD}M_{AC}.
\end{align*}
Let $J_\mu=\frac{M_{\mu4}}{\ell}, \mu=0,\cdots,3$, then
\begin{align*}
  [J_\mu,J_\nu]&=-\frac{M_{\mu\nu}}{\ell^2},\\
  [J_\mu,M_{\alpha\beta}]&=\eta_{\mu\alpha}J_\beta-\eta_{\mu\beta}J_\alpha,\\
  [M_{\mu\nu},M_{\alpha\beta}]&=\eta_{\mu\beta}M_{\nu\alpha}+\eta_{\nu\alpha}M_{\mu\beta}-\eta_{\mu\alpha}M_{\nu\beta}-\eta_{\nu\beta}M_{\mu\alpha}.
\end{align*}
These relations can be realized via  the following differential operators
 \begin{align*}
       J_\mu&=\partial_\mu+\frac{\eta_{\mu\alpha}x^\alpha x^\nu\partial_\nu}{\ell^2},\\
       M_{\mu\nu}&=\eta_{\mu\alpha}x^\alpha\partial_\nu-\eta_{\nu\alpha}x^\alpha\partial_\mu=\eta_{\mu\alpha}x^\alpha J_\nu-\eta_{\nu\alpha}x^\alpha J_\mu.
                                                                          \end{align*}
Let us introduce the following symbols
\begin{align*}
    K_\mathfrak{i}^{\pm}&=\frac{1}{\sqrt2} (M_{0i}\pm M_{1i}), \mathfrak{i}=2,3\\
    F_{i}^{\pm}&=\frac{1}{\sqrt2} (\frac{M_{0i}}{\ell}\pm J_{i}),i=1,2,3,\\
    L_{i}&=\frac{1}{2}\epsilon_{ijk}M_{jk}, i,j,k=1,2,3,\\P^{\pm}&= \frac{1}{\sqrt2} (J_0\pm J_1),\\
 R&=M_{01}, T=M_{23}.
                                   \end{align*}
The maximal Lie subalgebras of $\mathfrak{o}(1,4)$ are of 7 dimensions, which are exhibited in the following list
\begin{table}[ht]\caption{Maximal Subgroups  of $O(1,4)$}
\begin{tabular}{l|c|c}
  \hline
  & Generators & Algebraic Relations \\
    \hline
Type I & $\{K_2^{\pm},K_3^{\pm},J_2, J_3, P^{\pm},R,T\}$ & \begin{tabular}{c}
                                           $[K_\mathfrak{i}^{\pm},K_\mathfrak{j}^{\pm}] =0,[J_\mathfrak{i},J_\mathfrak{j}]=\epsilon_{\mathfrak{ij}}\frac{T}{\ell^2}, [K_\mathfrak{i}^{\pm},J_\mathfrak{j}]=\delta_{\mathfrak{ij}}P^{\pm}$,\\
  $[K_\mathfrak{i}^{\pm},P^{\pm}]=0,[K_\mathfrak{i}^{\pm},R]=-K_\mathfrak{i}^{\pm}, [K_\mathfrak{i}^{\pm},T]=\epsilon_{\mathfrak{ij}}K_\mathfrak{j}^{\pm}$,\\
  $[J_\mathfrak{i},P^{\pm}]=\frac{K_\mathfrak{i}^{\pm}}{\ell^2}, [J_\mathfrak{i},R]=0,[J_\mathfrak{i},T]=\epsilon_{\mathfrak{ij}}J_j$.\\
  $[P^{\pm},R]=\mp P^{\pm},[P^{\pm},T]=0,,[R,T]=0$,
                                           \end{tabular}
 \\ \hline
Type II & $\{F_1^{\pm},F_2^{\pm},F_3^{\pm},L_1,L_2,L_3,J_0\}$ & \begin{tabular}{c}
                                           $[F_{i}^{\pm},F_{j}^{\pm}]=0, [L_{i},L_{j}]=-\epsilon_{ijk}L_k, [F_{i}^{\pm},L_{j}]=-\epsilon_{ijk}F_k^{\pm}$,\\
                                           $[F_i^{\pm},J_0]=\pm\frac{F_i^{\pm}}{\ell^2},[L_i,J_0] = 0$.
                                           \end{tabular} \\
  \hline
\end{tabular}\end{table}

The little groups in $O(1,4)$ corresponding to these two types of Lie subalgebras are denoted by $\mathcal{G}$ and $\mathcal{H}$ respectively.  When the parameter $\ell$ tends to infinity, $\mathcal{G}$ are   subgroups of $ISIM(2)$, which are  8-dimensional maximal subgroups of the Poincar\'{e} group generated by $\{K_2^{\pm},K_3^{\pm},J_1, J_2, P^+,P^-,R,T\}$, and $\mathcal{H}$ are isomorphic to  the semiproduct of  $O(3)$ and 4-dimensional translation group $\mathbb{T}(4)$.
Note that there are no new invariant tensors for the groups $\mathcal{G}$ or $\mathcal{H}$, therefore we should consider the subgroups of $\mathcal{G}$ and $\mathcal{H}$.
\begin{example}Consider the subgroup $\mathcal{S}$ whose Lie algebra is generated by $K^+_2, K^+_3, P^+,T$.
By means of  the following matrix representations of generators
\begin{align*}
  K_2^+&=\frac{1}{\sqrt2}\left(
          \begin{array}{ccccc}
            0 & 0 & 1 & 0 & 0 \\
           0 & 0 & 1 & 0 & 0 \\
            1& -1 & 0 & 0 & 0 \\
           0 & 0 & 0 & 0 & 0 \\
           0 & 0 & 0 & 0 & 0 \\
          \end{array}
        \right), K_3^+=\frac{1}{\sqrt2}\left(
          \begin{array}{ccccc}
            0 & 0 & 0 & 1 & 0 \\
           0 & 0 & 0 & 1 & 0 \\
            0& 0 & 0 & 0 & 0 \\
           1 & -1 & 0 & 0 & 0 \\
           0 & 0 & 0 & 0 & 0 \\
          \end{array}
        \right),\\
        P^+&=\frac{1}{\ell}\left(
          \begin{array}{ccccc}
            0 & 0 & 0 & 0 & 1 \\
           0 & 0 & 0 & 0 & 1 \\
            0& 0 & 0 & 0 & 0 \\
           0 & 0 & 0 & 0 & 0 \\
           1& -1 & 0 & 0 & 0 \\
          \end{array}
        \right),T=\left(
          \begin{array}{ccccc}
            0 & 0 & 0 & 0 & 0 \\
           0 & 0 & 0 & 0 & 0 \\
            0& 0 & 0 & 1 & 0 \\
           0 & 0 & -1 & 0 & 0 \\
           0& 0 & 0 & 0 & 0 \\
          \end{array}
        \right),
\end{align*}
we find a second-order non-degenerate symmetric invariant tensor with respect to $\mathcal{H}$ 
\begin{align}
 C=\left(
     \begin{array}{ccccc}
       \textsf{ a} & \textsf{b} & 0 & 0 & 0 \\
           \textsf{b} & 2\textsf{b}-\textsf{a} & 0 & 0 & 0 \\
            0& 0 & \textsf{b}-\textsf{a }& 0 & 0 \\
           0 & 0 & 0 & \textsf{b}-\textsf{a} & 0 \\
           0& 0 & 0 & 0 & \textsf{b}-\textsf{a} \\
     \end{array}
   \right)
\end{align}
with two constants \textsf{a}$<$\textsf{b}, thus  a metric
\begin{align}
  C=\textsf{a}dT^2+2\textsf{b}dTdX+(2\textsf{b}-\textsf{a})dX^2+(\textsf{b}-\textsf{a})dY^2+(\textsf{b}-\textsf{a})dZ^2+b(\textsf{b}-\textsf{a})dW^2
  \end{align}
  Then the coordinate transformations \eqref{t} give rises to an induced metric
  \begin{align}
  C_{\mu\nu}dx^\mu dx^\nu =(\textsf{b}-\textsf{a})g_{\mu\nu}dx^\mu dx^\nu+\textsf{b}\frac{[(b+\eta_{\alpha\beta}x^\alpha x^\beta)(dx^0+dx^1)-(x^0+x^1)\eta_{\alpha\beta}x^\alpha dx^\beta]^2}{(b+\eta_{\alpha\beta}x^\alpha x^\beta)^3}.
  \end{align}
  Hence the $\mathcal{S}$-invariant action can be chosen as
\begin{align}
  S=-\int\sqrt{C_{\mu\nu} dx^\mu dx^\nu}.
\end{align}
  where the dimensionless constant $\textsf{b}$ is  set to be very small  characterizing the violation of the law of inertia.
\end{example}

\begin{example}
Consider the subgroup $\mathcal{V}$ whose Lie algebra is generated by $F^+_1, F^+_2, F_3^+$.
By means of  the following matrix representations of generators
\begin{align*}
  F_1^+&=\frac{1}{\sqrt2\ell}\left(
          \begin{array}{ccccc}
            0 & 1 & 0 & 0 & 0 \\
           1 & 0 & 0 & 0 & 1 \\
            0& 0 & 0 & 0 & 0 \\
           0 & 0 & 0 & 0 & 0 \\
           0 & -1 & 0 & 0 & 0 \\
          \end{array}
        \right), F_2^+=\frac{1}{\sqrt2\ell}\left(
          \begin{array}{ccccc}
            0 & 0 & 1 & 0 & 0 \\
           0 & 0 & 0 & 0 & 0 \\
            1& 0 & 0 & 0 & 1 \\
           0 & 0 & 0 & 0 & 0 \\
           0 & 0 & -1 & 0 & 0 \\
          \end{array}
        \right),
        F_3^+=\frac{1}{\sqrt2\ell}\left(
          \begin{array}{ccccc}
            0 & 0 & 0 & 1 & 0 \\
           0 & 0 & 0 & 0 & 0 \\
            0& 0 & 0 & 0 & 0 \\
           1 & 0 & 0 & 0 & 1 \\
           0&0 & 0 & -1 & 0 \\
          \end{array}
        \right),
\end{align*}
an invariant vector
\begin{align}
  V=(\textsf{a},0,0,0,-\textsf{a})^T
\end{align}
with a constant $\textsf{a}$.
Therefore we can consider a Finsler-type $\mathcal{V}$-invariant action \cite{gg,gg1}
\begin{align}
  S=\int(g_{\mu\nu}dx^\mu dx^\nu)^{\frac{1-\delta}{2}}(V_\mu dx^\mu)^\delta,
\end{align}
where 
\begin{align}
  V_0 &= \textsf{a}(b+\eta_{\mu\nu}x^\mu x^\nu)^{-\frac{3}{2}}(b+\bm{x}\cdot\bm{x}-x^0),\\
  V_i&= \textsf{a}(b+\eta_{\mu\nu}x^\mu x^\nu)^{-\frac{3}{2}}x^i(1-x^0),i=1,2,3,
\end{align}
and the dimensionless constant $\delta$ is  set to be very small  characterizing the violation of the law of inertia.
\end{example}
\section{Emergent Finsler   Geometry  }

We have seen that   our method in previous  sections cannot admit a universal constant  tending to zero. As mentioned in Introduction, we consider the bipartite-Finsler-like action \eqref{41}.
Here, we set $\xi=\frac{\kappa}{\ell}$ for a new universal constant $\kappa$ identified with Planck length.
Once again, one imposes the law of inertia on the corresponding Euler-Lagrange equation. Obviously, one can choose $B_{\mu\nu}$ to be the solution providing by \eqref{33}. Then the dynamical symmetry group breaks down to the Lorentz group $O(1,3)$.

  For simplicity, we  only work on the physical domain  $B-t^2+\bm{x}\cdot\bm{x}>0$ with $A>0, B>0$.
A Finsler metric  emerges from the action \eqref{41} as \cite{z2,z3}
\begin{align}
 g^{\textrm{F}}_{\mu\nu}:=&-\frac{1}{2}\frac{\partial^2\tilde L^2}{\partial v^\mu\partial v^\nu}\nonumber\\
=& -[\frac{\tilde L}{\lambda}\eta_{\mu\nu}+\xi\frac{\tilde L}{\sigma}B_{\mu\nu}+\xi \lambda\sigma k_\mu k_\nu],
\end{align}
where $k_\mu=\frac{\eta_{\mu\nu}v^\nu}{\lambda^2}-\frac{B_{\mu\nu}v^\nu}{\sigma^2}, \tilde L=\frac{L}{m}=-\lambda-\xi\sigma$ for $\lambda=\sqrt { -\eta_{\mu\nu} v^\mu v^\nu}$ and $\sigma=\sqrt { -B_{\mu\nu} v^\mu v^\nu}$.
Some dynamical effects can been studied via this Finsler metric.

  For example, we derive the new dispersion  relation for the massive free particle.
  We need to  calculate the inverse  $(g^{\textrm{F}})^{\mu\nu}$ of $g^{\textrm{F}}_{\mu\nu}$. In general, it is quite difficult. However, fortunately, for our case, taking advantage  of  \eqref{333}-\eqref{344},
we can explicitly obtain
\begin{align}
  (g^{\textrm{F}})^{\mu\nu}=&-\{(\frac{\tilde L}{\lambda}+\xi A\chi\frac{\tilde L}{\sigma})^{-1}\Phi^{\mu\nu} +(\frac{\tilde L}{\lambda}+\xi AB\chi^2\frac{\tilde L}{\sigma})^{-1}\Psi^{\mu\nu}\nonumber\\
  &\ \ \ \ \ -\frac{\xi \lambda\sigma}{1+\xi \lambda\sigma k^2}[(\frac{\tilde L}{\lambda}+\xi A\chi\frac{\tilde L}{\sigma})^{-1}\Phi^{\mu\alpha}+(\frac{\tilde L}{\lambda}+\xi AB\chi^2\frac{\tilde L}{\sigma})^{-1}\Psi^{\mu\alpha}]\nonumber\\
 &\ \ \ \ \ \ \ \ \ \cdot [(\frac{\tilde L}{\lambda}+\xi A\chi\frac{\tilde L}{\sigma})^{-1}\Phi^{\nu\beta}  +(\frac{\tilde L}{\lambda}+\xi AB\chi^2\frac{\tilde L}{\sigma})^{-1}\Psi^{\nu\beta}] k_\alpha k_\beta\},
\end{align}
where
 \begin{align*}
 \chi&=\frac{1}{B+\eta_{\mu\nu}x^\mu x^\nu},\\
   \Phi^{\mu\nu}&=\eta^{\mu\alpha}\eta^{\nu\beta}\Phi_{\alpha\beta},
   \Psi^{\mu\nu}=\eta^{\mu\alpha}\eta^{\nu\beta}\Psi_{\alpha\beta},\\
  k^2&= [(\frac{\tilde L}{\lambda}+\xi A\chi\frac{L}{\sigma})^{-1}\Phi^{\mu\nu}  +(\frac{\tilde L}{\lambda}+\xi AB\chi^2\frac{L}{\sigma})^{-1}\Psi^{\mu\nu}] k_\mu k_\nu.
 \end{align*}
  
Then the dispersion relation is given by
\begin{align}\label{mj}
  (g^{\textrm{F}})^{\mu\nu}\mathbb{P}_\mu \mathbb{P}_\nu=-m^2,
\end{align}
where 
\begin{align}
 \mathbb{P}_\mu=-\frac{\partial L}{\partial v^u}= \mathbb{P}^\eta_\mu+ \xi\mathbb{P}^B_\mu
\end{align}
 is the canonical 4-momentum in the sense of dynamics with 
 \begin{align*}
  \mathbb{P}^\eta_\mu=m\frac{\eta_{\mu\alpha}v^\alpha}{\lambda}, \ \mathbb{P}^B_\mu=m\frac{B_{\mu\alpha}v^\alpha}{\sigma}.
 \end{align*}
Expanding the identity \eqref{mj}   until the first order in $\xi$, we get
\begin{align}
  \lambda^2\eta^{\mu\alpha}\eta^{\nu\beta}(k_\alpha k_\beta-\frac{1}{\sigma^2}B_{\alpha\beta})\mathbb{P}^\eta_\mu\mathbb{P}^\eta_\nu+2\frac{\lambda}{\sigma}\eta^{\mu\nu}\mathbb{P}^\eta_\mu\mathbb{P}^B_\nu=-m^2.
\end{align}
More discussions on kinematics and dynamics in the general  bipartite-Finsler geometry can be found in \cite{z2,z3}.

\

\paragraph{\textbf{Acknowledgments}}The  the second-named author\footnote{It is very sadly that Prof. Mulin Yan passed away  after this manuscript had finished.}would like to thank  Prof. Ronggen Cai, Prof. Sen Hu, Prof. Hanying Guo, Prof. Si Li and Prof. Jianxin Lu for their useful discussions.

 \end{document}